\documentclass[lettersize,journal]{IEEEtran}
\usepackage{amsmath}                             
\usepackage{amsmath,amsfonts,amssymb,amsthm,cancel,siunitx,
calculator,calc,mathtools,empheq,latexsym}
\usepackage{hyperref}
\usepackage{subfig,epsfig,tikz,float}		            
\usepackage{booktabs,multicol,multirow,tabularx,array}   
\usepackage{algorithm}
\usepackage{algpseudocode}

\def\be{\begin{eqnarray}}
\def\ee{\end{eqnarray}}
\def\bee{\begin{eqnarray*}}
\def\eee{\end{eqnarray*}}

\def\E{\mathbb{E}}

\def\bmx{\begin{pmatrix}}
\def\emx{\end{pmatrix}}

\def\be{\begin{eqnarray}}
\def\ee{\end{eqnarray}}
\def\bee{\begin{eqnarray*}}
\def\eee{\end{eqnarray*}}

\newtheorem{thm}{Theorem}

\def\E{\mathbb{E}}

\def\bmx{\begin{pmatrix}}
\def\emx{\end{pmatrix}}
\def\bmat{\begin{pmatrix}}
\def\emat{\end{pmatrix}}

\def\Pf{\par\noindent{\em Proof:}\hskip0.1in}

\newcommand{\ignore}[1]{}

\title{ Feedback control for distributed ledgers: An attack mitigation policy for DAG-based DLTs}
\author{P. Ferraro\thanks{Pietro Ferraro and Robert Shorten are with Dyson School of Design Engineering, Imperial College London, London.}, A. Penzkofer\thanks{Andreas Penzkofer and Pietro Ferraro are with the IOTA Foundation, Berlin.}, C. King\thanks{Christopher King is with Department of Mathematics, Northeastern University, Boston.}, R. Shorten\thanks{All the code used to generate the figures in this work can be found at this  \href{https://github.com/V4p1d/Tips-Inflation-Attack}{link}.}
}

\begin{document}
\maketitle


\begin{abstract}
In this paper we present a feedback approach to the design of an attack mitigation policy for DAG-based Distributed Ledgers. We develop a model to analyse the behaviour of the ledger under the so called \emph{Tips Inflation Attack} and we design a control strategy to counteract this attack strategy. The efficacy of this approach is showcased through a theoretical analysis, in the form of two theorems about the stability properties of the ledger with and without the controller, and extensive Monte Carlo simulations of an agent-based model of the distributed ledger.

\end{abstract}
\baselineskip=\normalbaselineskip

\section{Introduction}\label{sec:1}
A basic assumption in the design and operation of distributed ledgers (DLTs) is that they will operate in highly adversarial environments that are characterised by large numbers of untrustworthy agents. Indeed much of the recent literature on this topic is concerned with developing DLTs that are robust in the presence of such actors. In addition to a large body of work on identifying attack scenarios and vulnerabilities in ledger architectures, roughly speaking, work on attack mitigation has developed along three main directions. One approach, based on game theory is to design incentive mechanisms for ledgers that make certain attacks prohibitively expensive for attackers, so that the resulting protocol is characterised by a Nash equilibrium \cite{kiayias2016blockchain}. A second approach is based on developing techniques to identify specific attacks and to make ad-hoc modifications to the protocol \cite{cullen2020resilience}. Third, machine learning techniques can be utilised to map the spectrum of attack vulnerabilities associated with a ledger structure and to make modifications accordingly in an iterative manner. Often these approaches affect the basic operation of the ledger structure in order to eliminate attack vulnerabilities.\newline 

We propose a new direction in this work in which concepts from control theory can be used to develop policies that protect ledgers against certain adversarial attacks.  The rationale for developing such strategies is that they preserve the operational efficiencies of the basic ledger in honest environments, but also provide protection when needed (potentially at the cost of some operational efficiency during periods of attack). 
{The protection mechanism draws from the capability of nodes to agree, with high accuracy, on some parameters of the ledger. This knowledge is then utilised by the nodes to act in a coordinated way to counter the effects of the attack.}

The attack vector that motivates this work is the so called {\em Tips Inflation Attack} that is of concern in certain Directed Acyclic Graph (DAG) based DLTs. DAG based ledgers have emerged in recent years as a generalisation and alternative to blockchains, in which the ledger structure evolves not as a sequence of blocks, but rather graph like 
\cite{Wang2020SoKDI}. The utilisation of a DAG structure allows for several interesting features, such as faster block writing times and partial ordering, all of which enable interesting use cases of DLTs \cite{overko2020spatial, katsikouli2020distributed, ferraro2021personalised}. 
DAGs, however, raise several concerns with regards to the vulnerability of the data structure that are not present in classical blockchains. For example in chain-based systems, the growth of the data structure is limited by the block issuance rate and  blocks that do not extend the longest chain are removed or   ``orphaned''. In contrast, in DAG-based systems the extension of the data structure is less clear and depends significantly on the protocol that is adopted for the validation of new blocks.\footnote{For example, in \cite{popov2015} the author proposes only to extend the data structure at places that ``support'' the preferred state of the ledger.} This enables new attack vectors to the data structure itself that must be studied and, where possible, averted.  This paper describes an attack on a  {well known DAG-based DLT, namely IOTA.} In this attack the adversary attempts to modify the block validation process in  {the block-DAG, which is also called \textit{Tangle}, }such that the number of {\em orphaned} blocks grows unbounded. 
The basic idea is that at some time a malicious agent that obtains a proportion $q \in [0,1]$ of the writing access to the DAG, starts publishing blocks that approve only own and old blocks. This, as we will see, results in an increase to the number of unconfirmed blocks or    ``\textit{tips}'' in the Tangle, potentially leading to an instability of the system.\newline

\noindent Specifically, the contributions of this paper are as follows.
\begin{itemize}
    \item We perform a theoretical analysis of the Tips Inflation Attack. Sufficient conditions for its success are provided in the form of a stability theorem for the Tangle.
    \item A modification of the IOTA protocol, inspired by control theory. The efficacy of this approach is shown through theoretical results and extensive simulations.
\end{itemize}

 We note that this work builds on top of previous research \cite{ferraro2018distributed,ferraro2019stability}, in which the authors derived a fluid model for the Tangle \cite{ferraro2018distributed} and proposed an alternative tip selection algorithm \cite{ferraro2019stability} to ensure the stability of the network.

\section{{Structure of the paper}}

The remainder of this paper is organised as follows. Sections \ref{sec: Tangle} and \ref{sec: attack} present the original version of the IOTA protocol and the Tips Inflation Attack. In Section \ref{sec: model} we present the mathematical model used to analyse the attack. In Section \ref{sec: control algorithm} we introduce a modification of the tip selection algorithm that offers a solution to the attack. in Section \ref{sec: Simulations} we validate the theoretical results through extensive Monte Carlo simulations and finally, in Section \ref{sec: Conclusions} we summarise our findings and present future lines of research.

\section{The Tangle}
\label{sec: Tangle}
We are interested in a particular DLT architecture that makes use of a Directed Acyclic Graph (DAG) $D=(V,E)$, with vertex set $V$ and directed edges set $E$, to achieve consensus on a shared ledger. 
A particular instance of a DAG based DLT is introduced in \cite{popov2015}. In this DAG, called \textit{Tangle}, the vertices are blocks issued by network participants, called \textit{nodes}, and each block contains at most one transaction.\footnote{ {This limitation is not a necessity and the number of transactions per block can, in principle, be increased.}} The edges are formed by blocks approving previous blocks that we call \textit{parents}. Since the approval can be done only toward past blocks, the graph is acyclic, starting with a first block, called the \textit{genesis}. All yet unapproved blocks are called \emph{tips} and the set  of all unapproved blocks is called the \emph{tips set}. A node selects tips to approve through a \textit{tip selection algorithm}.
To issue a block a node selects $k$ blocks from the tips set.\footnote{Nodes are not necessarily obliged to select a given number of parents. But it is reasonable to expect that most of them will follow the suggested number.}
This process, called \textit{approval}, is represented by an edge in the graph. If there exists a directed edge from vertex $i$ to $j$, we say $j$ is \textit{directly approved} by $i$. If there is a directed path from $i$ to $j$ we say that  $j$ is \textit{indirectly approved} by $i$.

Due to various factors, there is a delay between the selection of the parents of a block and their removal from the tip pool. First, there may be delays at the issuing node due to  Proof of Work (PoW). The role of PoW is to prevent malicious users from spamming the network, thus, the required PoW is less computationally intense than in blockchain counterparts  \cite{karame2016bitcoin, gervais2016security}, and can be easily carried out by common IoT devices, e.g. smartphones and smart appliances. Second, the time until the parents are removed in the tip pools of other nodes is delayed by propagation times and processing. As a consequence it is possible that a block is approved several times by multiple other blocks. 
This also leads to each node having an own local version of the Tangle. 
We call the period between selection of the parents and the appearance in the tip pool, the \textit{delay period}.

\begin{figure*}[h]
\centering
\includegraphics[width=2\columnwidth]{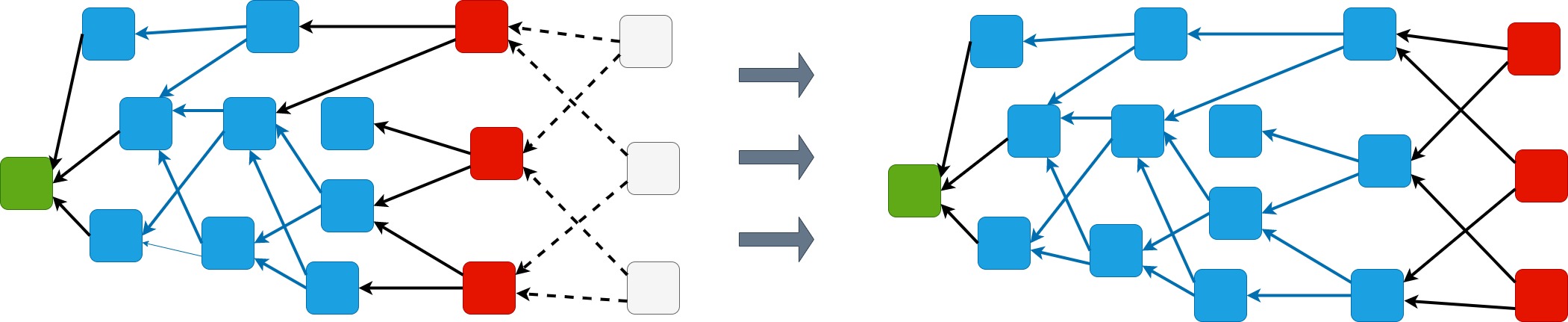}
\caption{Sequence to issue a new block. The blue sites represent the approved blocks, the red ones represent the tips and the gray ones represent newly arriving blocks. The black edges represent approvals, whereas the dashed ones represent edges  {from blocks that are issued, but are not yet visible. After the grey blocks become visible the red blocks cease to be tips.}}
\label{Fig: Tangle}
\end{figure*}

{As an example of this process, Figure \ref{Fig: Tangle} shows an instance of the Tangle with three new incoming blocks (left panel). Blue blocks have already been approved, red blocks represent the current tips of the Tangle and grey blocks are incoming, i.e. are in the delay period. 
Notice that at this stage, the newly arriving tips remain unconfirmed (dashed lines) until this process is over. Once the delay period is over, the selected tips become confirmed vertices and the grey blocks are added to the tips set (right panel).}

\subsection{Tip selection algorithm}

{In the IOTA protocol the participants in the network, i.e. nodes, are the creators of blocks. In order to extend the Tangle and advance the confirmation of blocks, nodes are required to select previous blocks and reference these \cite{Kusmierz2019}.}

In \cite{popov2015} new nodes selects tips as on the basis of $k$ weighted random walks, from the interior of the DAG to the tips set. It can be proven that this protocol represents a Nash equilibrium for the Tangle \cite{popov2019equilibria}, which each node is incentivised to follow, as any deviation from this tip selection algorithm results into orphanage of the own blocks. 
We note the algorithm protects against an adversary in the presence of an honest majority of issuers. However, the computational cost of running the random walk algorithm in the graph  is prohibitive in an IoT scenario.

{A more computationally efficient algorithm for selecting parent blocks is the \textit{uniform random tip selection} (URTS), in which each tip has the same probability of being selected by a new block. However, in order to prevent penalisation of block issuers for how they append, it must be ensured that the attachment location has no or little impact on the liveness of the blocks.
This can be {, for example,} achieved by operating a consensus mechanism prior to adding blocks to the Tangle, e.g. \cite{2020Muller}, which ensures that no conflicts enter the tip set.
}

\section{Tips Inflation attack}\label{sec: attack}

{In the previous section we introduced the Tangle and how it can be constructed using a URTS algorithm. Under honest conditions this leads to a stable tip pool size. We now describe an attack that attempts to increase the tip pool size, potentially unbounded. }

\begin{figure*}[h]
\centering
\includegraphics[width=2\columnwidth]{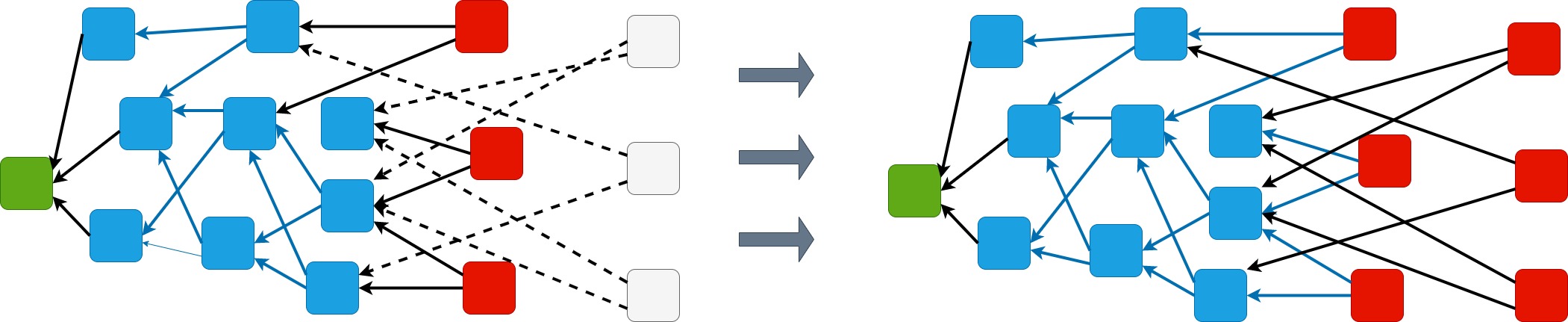}
\caption{Example of the Tips Inflation Attack, in the extreme case in which all the new blocks do not approve tips. The amount of tips keeps increasing over time.}
\label{fig:attack}
\end{figure*}

{Typically blocks in the Tangle reference more than one parent in order to keep the tip pool size bounded. 
In the attack an adversary modifies the tip selection algorithm, such that the adversary's blocks only approve old blocks (which are not tips). This may be, for example, the genesis or alternatively "old" blocks. If an attacker manages to gain a  {proportion} $q$ of the writing power of the network, then for sufficiently large $q$, the attacker can artificially increase the size of the tips set arbitrarily, provided that the attack is maintained.\footnote{The writing power $q$ depends on the specific protocol that the network is following: it can be the computing power of the attacker with respect to the rest of the network or some weight \cite{cullen2021} that is associated with a specific node. For the purpose of this work, the means through which an attacker gets writing power $q$ is irrelevant.} Figure \ref{fig:attack} shows an example of this process. This would lead to liveness problems, where honest blocks are not approved for an extensive amount of time or, in the worst case not approved at all. }

One may consider, as a basic protection, to only allow nodes to approve blocks with a certain maximum age, i.e. blocks that are issued within a certain recent time $\Delta$. However, this would not mitigate the problem, as the attacker would be effectively decreasing the chance of honest block to be selected, regardless. What is more, since this introduces an \textit{expiration time} this may lead to orphanage of blocks, including the honest ones. 
Accordingly, in this paper we consider a version of the protocol with no constraints on the maximum allowed age $\Delta$.

\section{Model}
\label{sec: model}


{As mentioned in the previous section, there is a delay period between the selection of the parents for a block and the appearance of the block in the tip pool. Generally, the time of appearance for a given block is different  for each node, i.e. each node has an own local version of the Tangle.  However, for simplicity, we assume that each node has instantaneous reading access to a ``global'' Tangle, while the writing access is delayed by $h$. Thus, during a time period $h$ the approvals of the selected tips are pending and tips may still be available for selection by other new blocks, although the block has already been selected by a node. After the delay time $h$ the new directed edges are added to the global graph, directed from the new site to its parent sites. After this point the parent sites are no longer tips, and thus are no longer available for selection by other new blocks.\footnote{ {If a block $z$ is referenced by two blocks $x$ and $y$, at times $t_x$ and $t_y$ respectively, where $t_x<t_y$, then block $z$ ceases to be a tip at $t_x+h$.}} In an honest environment, the URTS leads to a stabilisation of the global Tangle tip pool size at the value $L=\frac{k}{k-1}\lambda h$, where  $\lambda$ is the block issuance rate \cite{Kusmierz2019}.}

\medskip
\noindent Our model involves the following variables:
\begin{itemize}
\item[1)] $L(t)$ is the number of tips at time $t$
\item[2)] $W(t)$ is the number of `pending' tips at time $t$ which are being considered for approval by some new block(s)
\item[3)] $X(t) = L(t) - W(t)$ is the number of `free' tips at time $t$
\item[4)] $T_a$ is the time when block $a$ is created
\item[5)] $N(t)$ is the number of blocks created up to time $t$
\item[6)] $U(T_a) \in \{0,1,2\}$ is the number of free tips selected for approval by block $a$ at time $T_a$
\end{itemize}

We have the relations
\be\label{eqn:N}
N(t) &=&  \sum_{a : T_a \le t} \\
W(t) &=& \sum_{a: t-h < T_a \le t} U(T_a) \label{eqn:W} \\
X(t) &=& N(t -h) - \sum_{a : T_a \le t} U(T_a) \label{eqn:X} \\
L(t) &=& N(t-h) - \sum_{a : T_a \le t - h} U(T_a) \label{eqn:L}
\ee

\medskip
Assuming the uniform random tip selection algorithm,
$U(T_a)$ is a random variable whose distribution depends only on the values
of $X$, $W$ and $L$ at time $T_a$. (For convenience we assume that block $a$ is created at time
$T_a + 0$, so it sees the state of the Tangle at time $T_a$).
The expected value of $U(T_a)$ is
\be\label{av-Ui}
\E[U(T_a)] = 2 \, \frac{X(T_a)}{L(T_a)} -  \frac{X(T_a)}{L(T_a)^2}
\ee

\subsection{Fluid Model for two selections}

In order to gain some understanding of the system of equations (\ref{eqn:N})-(\ref{eqn:L}) we consider the asymptotic regime of large arrival rate,
where the time between consecutive blocks is very small. In this regime
it should be reasonable to approximate the system  (\ref{eqn:N})-(\ref{eqn:L}) by a fluid model.

We introduce a scaling parameter $\lambda$ so that the arrival rate is proportional to $\lambda$,
and let $\lambda \rightarrow \infty$ to reach the fluid model. The
rescaled variables $\{\lambda^{-1} L(t), \lambda^{-1} X(t), \lambda^{-1} W(t)\}$ are assumed to converge
to deterministic limits as $\lambda \rightarrow \infty$, and the limits are represented
in the fluid model by real-valued functions $\{l(t),x(t),w(t)\}$.
The creation of new blocks in the fluid model is described by an arrival rate $a(t)$ so that
$\int_{0}^t a(s) \, ds$ corresponds to the limit of $\lambda^{-1} N(t)$. 
Furthermore the variable
$U(T_a)$ is replaced by its time average (over a short time interval), which
by the law of large numbers  is equivalent to the ensemble average (\ref{av-Ui}).
By rescaling variables and letting $\lambda \rightarrow \infty$, this expected value converges to
$2 x(t)/l(t)$.
Referring to the equation (\ref{eqn:X}) the change of $X$ over a small time increment $\gamma$
can be approximated as
\bee
&& \hskip-0.4in X(t + \gamma) - X(t)  \\
&=& N(t -h + \gamma) - N(t - h) -
\sum_{a : t< T_a \le t+\gamma} U(T_a) \\
& \simeq & \gamma \, \lambda\, a(t - h) \, - \gamma \, \lambda\, a(t) \,  \frac{2 x(t)}{l(t)} \\
& = & \gamma \, \lambda \left[a(t - h) \, - a(t) \, \frac{2 x(t)}{l(t)} \right]
\eee
Applying similar reasoning to the other equations we get
the following set of delay differential equations (DDE) for the fluid model:
\be
\frac{d x}{d t}(t) &=& a(t - h)-  a(t) \, u(t) \label{DDEx} \\
\frac{d l}{d t}(t) &=& a(t - h) -  a(t - h) \, u(t-h) \label{DDEl} \\
w(t) &=&  l(t) - x(t) = \int_{t-h}^t a(s) \, u(s) d s  \label{DDEw}
\ee
where
\be\label{DDE2}
u(t) =   \frac{2 \, x(t)}{l(t)}
\ee
The solution of these equations $\{x(t), l(t), w(t)\}$ can be interpreted as a fluid model which
describes the dynamics of the Tangle with very high arrival rate, using the random tip selection
algorithm. Note that
the DDE system (\ref{DDEx}) - (\ref{DDEw}) must be supplemented by initial conditions in the interval $[-h,0]$.

\subsection{Fluid Model for k selections and Tips Inflation Attack}

In the presence of a Tips Inflation attack, a malicious actor will attach a proportion $q$ of the new tips to blocks that have already been referenced, effectively reducing the number of tips that are removed from the tips set.
Accordingly, using the same derivation outlined in the previous section, we can derive a fluid model that describes the dynamics of the Tangle with very high arrival rate, using a random tip selection algorithm but now with $k$ parents:

\be
\frac{d x}{d t}(t) &=& a(t - h)-  p\, a(t) \, u(t) \label{DDExq} \\
\frac{d l}{d t}(t) &=& a(t - h) -  p\, a(t - h) \, u(t-h) \label{DDElq} \\
w(t) &=&  l(t) - x(t) = \int_{t-h}^t p \, a(s) \, u(s) d s  \label{DDEwq}
\ee
where
\be\label{DDE2q}
u(t) =   \frac{k \, x(t)}{l(t)}
\ee

Considering this new model, for a constant arrival rate $a(t) = a(t-h) = \beta$, it is straightforward to derive the equilibrium $(\hat{x}, \hat{l})$ of the system at steady state. From relations \ref{DDExq}-\ref{DDElq} we obtain:

\be
0 = 1 - pk \frac{ \hat{x}}{\hat{l}},
\ee
while from relation (\ref{DDEwq}) we obtain:
\be
\hat{l} - \hat{x} = \int_{t-h}^t p\, \beta \, k\frac{ \hat{x}}{\hat{l}} d s  =
 p \,\beta k\frac{ \hat{x}}{\hat{l}} (t - t + h) = \beta\,  p \, k\, h \,\frac{ \hat{x}}{\hat{l}}
\ee
These lead to 
\be
\hat{x} = \frac{h\beta}{pk - 1}
\ee
\be
\hat{l} =\frac{pkh}{pk - 1}\beta
\ee
\be
\hat{w} =\frac{pkh - h}{pk - 1}\beta
\ee
Notice that for $p = 1$ (corresponding to no attack) and $k = 2$ the equilibrium for the tips is $\hat{l} = 2h\beta$.

Notice also, that for $p \rightarrow \frac{1}{k}$, $\hat{l} \rightarrow \infty$ and for all $(p, k)$ such that $pk-1 <0$, $\hat{l} < 0$.

The previous observations lead to the following result:

\begin{thm}\label{thm1}
Consider the system  (\ref{DDExq} - \ref{DDE2q}) with constant arrival rate   $a(t)=\beta$.
For each pair $(p,k)$, $p \in [0,1]$, $k > 0$, the system has a stable solution $\{\hat{x}, \hat{l}, \hat{w}\}$, if $pk-1 >0$.
\end{thm}

\medskip
\proof Set $pk = c \neq 1$. In order to investigate the stability of the equilibrium $(\hat{x}, \hat{l}, \hat{w})$ we linearize equations \ref{DDExq}-\ref{DDElq} around $\{\hat{x}, \hat{l}, \hat{w}\}$. To do so, consider a small perturbation $(\zeta(t), \theta(t), \eta(t))$. Let $x_h(t) = x(t-h)$ and $y_h(t) = y(t-h)$. At linear order:
\be
\begin{bmatrix}
\dot{(x + \zeta)} \\
\dot{(l + \theta)}
\end{bmatrix}   &=&
\begin{bmatrix}
f_1(x(t),l(t))\\
f_2(x_h(t),l_h(t))
\end{bmatrix} + \\
& & +\begin{bmatrix}
\dfrac{df_1}{dx}|_{(\hat{x}, \hat{l}, \hat{w})} \zeta + \dfrac{df_1}{dl}|_{(\hat{x}, \hat{l}, \hat{w})} \theta \nonumber \\
\dfrac{df_2}{dx_h}|_{(\hat{x}, \hat{l}, \hat{w})} \zeta_h + \dfrac{df_2}{dl_h}|_{(\hat{x}, \hat{l}, \hat{w})} \theta_h
\end{bmatrix}
\ee
This leads to 
\begin{equation}
\begin{bmatrix}
h\dot{\zeta} \\
h\dot{\theta}
\end{bmatrix}   =
\begin{bmatrix}
-(c-1)\zeta + \dfrac{c-1}{c}\theta \\
-(c-1)\zeta_h + \dfrac{c-1}{c}\theta_h
\end{bmatrix}.
\end{equation}
The solution $\{\hat{l},\hat{x},\hat{w}\}$ is locally stable if 
\be
\max \{|\zeta(t)|, |\theta(t)|, |\eta(t)| \} \rightarrow 0 \quad \text{as $t \rightarrow \infty$}
\ee
for all solutions of the linear system. The solution is locally unstable if there is some solution
satisfying ${\cal L}^{(0)}(\zeta, \theta, \eta)=0$ such that
\be
\max \{|\zeta(t)|, |\theta(t)|, |\eta(t)| \} \rightarrow \infty \quad \text{as $t \rightarrow \infty$}
\ee
The spectrum of the linear system ${\cal L}^{(0)}(\zeta, \theta, \eta)=0$ is the set of complex values
$z$ for which the system has a nonzero solution of the form
\be
(\zeta(t), \theta(t), \eta(t)) = e^{z t} \, (\zeta, \theta, \eta)
\ee
for some constants $(\zeta, \theta, \eta)$. The solution is stable if the spectrum is contained in the open
left half of the complex plane, and is unstable if there is some element of the spectrum in the open right half of the
complex plane.

Accordingly, consider a perturbation of the form
\be\label{loc-stab2}
\zeta(t) = \zeta e^{zt}, \quad \theta(t) = \theta e^{zt}.
\ee
This leads to the equations 

\begin{equation}
\begin{bmatrix}
\zeta hz\\
\theta hz e^{hz}
\end{bmatrix}   =
\begin{bmatrix}
-(c-1)\zeta + \dfrac{c-1}{c}\theta \\
-(c-1)\zeta + \dfrac{c-1}{c}\theta
\end{bmatrix}.
\end{equation}

Algebraic manipulation of the expressions above leads to the equation
\be
c + \dfrac{c}{c-1}hz = e^{-hz}
\ee
It can be verified that this equation has no roots in the open right half plane for all $c >1$ and has roots in the open right half plane for  all $ c \in [0,1)$, which implies that the solutions $\{\hat{x}, \hat{l}, \hat{w}\}$ are locally unstable for $c \in [0,1)$ and locally stable for $c > 1$. This concludes the proof.
\subsection{Validation of the previous results}
\label{sec: validation}
\begin{figure}
\includegraphics[width=\columnwidth]{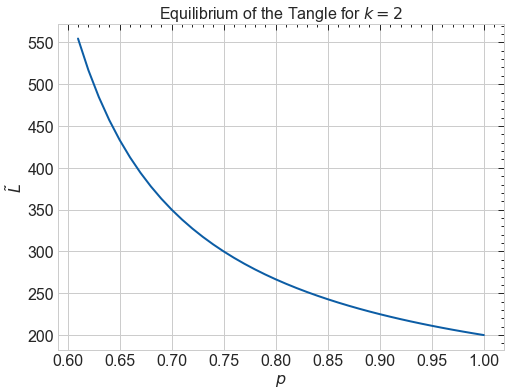}
\caption{Equilibria for the tip pool size in the Tangle vs percentage of honest new blocks. The values of $\Tilde{L}$ are obtained considering $\beta = 100$ and $h = 1$.}
\label{Fig: Comparison 1}
\end{figure}

\begin{figure}
\includegraphics[width=\columnwidth]{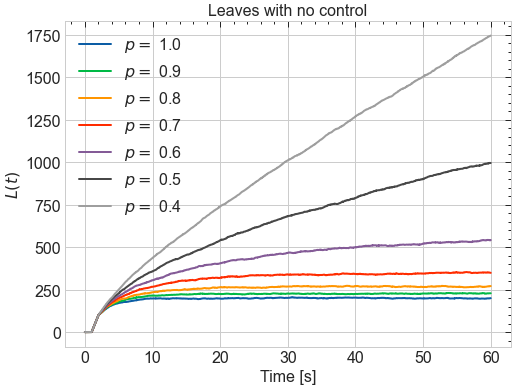}
\caption{Simulations of the Tangle for different values of $p$ and $k=2$. Each simulation shows 100 realizations of the model with block times, $T_a$, generated according to a Poisson distribution with $\beta = 100$ and $h = 1$. Notice that for $p = 0.5$ and $p = 0.4$ the system becomes unstable, while for all other values of $p$ the average steady state value of the system reaches the equilibrium predicted by the fluid model. }
\label{Fig: Comparison 2}
\end{figure}

To validate the model and the result presented in the previous paragraph we compare its behaviour with an agent based version of the Tangle: at each time step a random number of blocks arrive, according to a Poisson distribution, and for each one of these blocks the tip selection algorithm is performed on the current tips set in order to generate graph structures equivalent to the ones presented in detail in Section III. In other words, this agent based model simulates the behaviour of each block, therefore providing an accurate replica of the mechanism described in Section III. To simulate a Tips Inflation Attack, we assume that each new block will select two tips with  {proportion $p$ (representing the honest issuers) or no tips with proportion $q=1-p$ (representing the adversary).\footnote{Technically the latter is the case, where all referenced blocks are no longer tips.}}
The variables used for the comparison are the number of leaves $L(t)$, which are obtained by enumerating the number of tips at the end of each time step. Due to the stochastic nature of the Tangle, 100 Monte Carlo simulations are performed in order to obtain statistically meaningful results. Figures \ref{Fig: Comparison 1} and \ref{Fig: Comparison 2} show, respectively, the equilibria of the fluid model for $k=2$, as $p$ varies, and the corresponding agent based simulation for 6 different values of $p$. It is easy to see, by visual inspection, that the values and the stability behaviour predicted by the fluid model are also exhibited by the average steady state of the agent-based system\footnote{As a side note, we want to point out that the same conclusions appear to hold  also for scenarios with low arrival rates in which the fluid limit approximation would not hold. This has been observed through a large number of Monte Carlo simulations that can be found at this \href{https://github.com/V4p1d/Tips-Inflation-Attack}{link}.}.

\section{Control algorithm}
\label{sec: control algorithm}

Given the previous result, we propose a control algorithm  solution to tackle the Tips Inflation Attack. The approach is to add a feedback loop on the system that regulates the number of tips $k$ that each new block selects.

\be
k(t) = 2 + \dfrac{m}{\beta h}(l(t) - 2\beta h)^+
\label{eq: control}
\ee

Combining (\ref{eq: control}) and (\ref{DDExq})-(\ref{DDE2q}), under the assumption of a constant arrival rate $a(t) = \beta$, we obtain the following system of equations:

\be
\frac{d x}{d t}(t) &=& \beta( 1 -  p \, k(t) \frac{x(t)}{l(t)}) \label{DDExc} \\
\frac{d l}{d t}(t) &=& \beta( 1 -  p \, k(t-h) \frac{x(t-h)}{l(t-h)}) \label{DDElc} \\
w(t) &=&  l(t) - x(t) = \int_{t-h}^t p \, \beta \, k(s) \frac{x(s)}{l(s)} d s  \label{DDEwc} \\ 
k(t) &=& \dfrac{m}{\beta h}(l(t) - 2\beta h) \label{DDEc}.
\ee

For $p=1$ the equilibrium occurs at $k=2$, with $\Tilde{l} = 2 \beta h$ and  $\Tilde{x} = \beta h$. For $p < 1$ the equilibrium
occurs at $\Tilde{k} > 2$, with
\be
\Tilde{l} = K \frac{\beta h}{2 pm}, \hspace{5 mm}\Tilde{x} = K \frac{\beta h}{pm (4p - 4 pm + K )}
\label{eq: contrEquilibria}
\ee
and 
\be
\Tilde{k} = \frac{K}{2p} + 2 - 2 m,
\ee
where
\be
K = 3pm + 1 - 2p + \sqrt{(2p -1 -pm)^2 + 4pm}
\ee

\emph{Remark: } Notice that control law (\ref{eq: control}) implies the simplifying assumption that the arrival rate $\lambda$ is constant. While this assumption might look arbitrary, the Tangle is designed to work at capacity all the time (i.e. the amount of blocks that are added to the Tangle is the maximum allowed by the bandwidth of the system) and therefore assuming that the arrival rate remains constant, or varies slowly with time, is a reasonable assumption that would hold true in the majority of scenarios.

\subsection{Stability of the proposed control law}
Without loss of generality we take $\beta=1$ and consider the following more general control law:
\be
k(t) = 2 + r \, \left(l(t) - C\right)^{+} \label{eqs4}.
\ee
Notice that (\ref{eq: control}) is a special case of (\ref{eqs4}). Under these assumptions, the model depends on four parameters $r,C,h,p$, where $r,C,h > 0$ and $0 < p \le 1$.
The equations are
\be
\frac{d}{dt} x(t) &=& 1 - p \, k(t) \, \frac{x(t)}{l(t)} \label{eqs1} \\
\frac{d}{dt} l(t) &=& 1 - p \, k(t-h) \, \frac{x(t-h)}{l(t-h)} \label{eqs2} \\
w(t) &=& p\, \int_{t-h}^t k(s) \, \frac{x(s)}{l(s)} \, d s \label{eqs3} 
\ee
From (\ref{eqs1}-\ref{eqs3}) we get
\be\label{init}
w(t) = h - x(t) + x(t - h), \quad \frac{d}{dt} \left(l(t) - x(t) - w(t)\right) = 0.\nonumber
\ee
We will assume continuous initial conditions for $x,w,l$ on the interval $[-h,0]$, satisfying
\be\label{init2}
x(t), w(t), l(t) \ge 0, \quad l(t) = x(t) + w(t)\, \forall \, t \in [-h, 0].
\ee
With these conditions, it follows that for all $t \ge 0$ the values of $l(t), w(t)$ are fully determined by the values $(x(t), x(t-h))$,
that $l(t) = x(t-h) + h$,
and that the function $x(t)$ satisfies the following delay differential equation:
\be\label{DDE}
\frac{d}{dt} x(t) = 1 - p \, \frac{\left[2 + r \, (x(t - h) + h - C)^{+})\right] \,x(t)}{x(t-h)+h} 
\ee
The method of steps can be used to show that the equation (\ref{DDE}), together with initial conditions satisfying (\ref{init2}),
has a unique positive continuous solution for all $t \ge 0$. The equations
$w(t) =  h - x(t) + x(t - h)$ and $l(t) = x(t) + w(t)$ then provide the unique solution of the system
(\ref{eqs1}-\ref{eqs3}) for all $t \ge 0$.

\begin{thm}
Suppose that $x,w,l$ satisfy the system (\ref{eqs1}-\ref{eqs3}) with initial conditions on
$[-h,0]$ satisfying (\ref{init2}). Suppose also that $l(t) \le C$ for all $t \in [-h,0]$. Then
for all $T \ge 0$,
\be\label{thm:eq1}
l(T) \le h + \max\left[C, \frac{C}{2 p}, \frac{1}{r \, p}\right]
\ee
\end{thm}

\Pf
If $l(T) \le C + h$ then the bound (\ref{thm:eq1}) holds. Suppose $l(T) > C + h$, and define
\bee
S = \sup\left\{s \le T: l(s) \le C\right\}
\eee
Since $l(0) \le C$, it follows that $S \ge 0$, and continuity of $l$ implies that $l(S) =C$.
Clearly 
\be\label{l-bd0}
l(t) \ge C \quad \text{for all $t \in [S,T]$}.
\ee
Equation (\ref{eqs2}) implies that
\be\label{l-bd1}
l(t+h) = l(t) + \int_{t}^{t+h} l'(s) \, ds \le l(t) + h
\ee
It follows that for all
$t \in [S,S+h]$ we have $l(t) \le l(S) + h \le C + h$, and therefore $T \ge S + h$.
Define
\bee
K_1 &=& \min\left(r, \frac{2}{C}\right) \\
K_2 &=& \max\left(C, \frac{1}{p \, K_1}\right)
\eee
For all $t \in [S,T]$,
\bee
\frac{k(t)}{l(t)} &=& \frac{2 - r \,C}{l(t)} + r \\
& \ge & \begin{cases} r & \text{if $r\, C \le 2$} \cr
2 \, C^{-1} & \text{if $r\, C \ge 2$} \quad \text{(using the bound (\ref{l-bd0}))} \end{cases} \\
& \ge & K_1
\eee
This implies
\bee
\frac{d}{dt} x(t) \le 1 - p \, K_1 \, x(t) \quad \text{for all $t \in [S,T]$},
\eee
and hence for all $t \in [S,T]$
\bee
x(t) \le \left(x(S) - \frac{1}{p \, K_1}\right) \, e^{- p \, K_1 \, (t-S)} + \frac{1}{p \, K_1}.
\eee
Since $x(S) \le l(S) = C$,
\bee
x(t) & \le & \left(C - \frac{1}{p \, K_1}\right) \, e^{- p \, K_1 \, (t - S)} + \frac{1}{p \, K_1} \\
& \le & \begin{cases} \frac{1}{p \, K_1} & \text{if $C \le \frac{1}{p \, K_1}$} \cr
C & \text{if $C \ge \frac{1}{p \, K_1}$} \end{cases}
\eee
Therefore for all $t \in [S,T]$ we have
\be\label{x-bd1}
x(t) \le \max\left(C, \frac{1}{p \, K_1}\right) = K_2
\ee
Using $l(T) = x(T) + w(T) = x(T-h) + h$, and $T- h \ge S$, we deduce that
\bee
l(T)  \le  K_2 + h 
& = & \max\left(C, \frac{1}{p}  \max\left(\frac{1}{r}, \frac{C}{2}\right)\right) + h \\
& = & \max\left[C, \frac{1}{r p}, \frac{C}{2 \, p}\right] + h
\eee
which completes the proof.

\begin{algorithm}
\caption{Tips Management Control}
\label{alg: TMC}
\begin{algorithmic}[1]
\State For each new block $a$
\State Compute  $k(t)$
\State Compute $ \bar{k} = \lfloor k(t) \rfloor, t = k(t) - \bar{k})$;
\State Sample a random number $x$ from the uniform distribution in $[0,1]$ 
\If{$x \leq t$} 
    \State Block $a$ selects $\bar{k}$ tips
\Else
    \State Block $a$ selects $\bar{k} + 1$ tips
\EndIf 
\end{algorithmic}
\end{algorithm}

\section{Implementation Details and Simulations}
\label{sec: Simulations}
The control law (\ref{eq: control}) is not implementable directly on the Tangle as each block needs to select an integer number of blocks. To circumvent this issue, consider Algorithm \ref{alg: TMC}. It is easy to show that this stochastic behaviour converges in the fluid limit to control law (\ref{eq: control}). 

\begin{figure}
\includegraphics[width=\columnwidth]{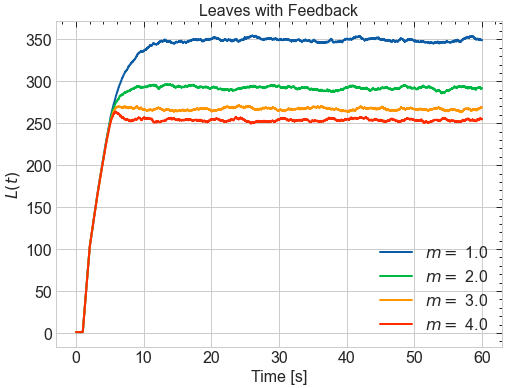}
\includegraphics[width=\columnwidth]{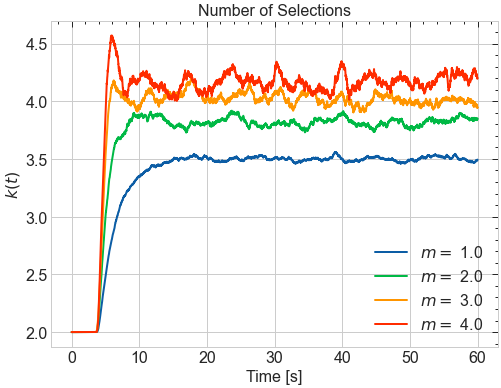}
\caption{The middle and bottom panels show simulations of the Tangle with Algorithm \ref{alg: TMC} for different values of $m$ and $p = 0.4$. Each simulation shows the average of 100 realizations of the model with block times, $T_a$, generated according to a Poisson distribution with $\beta = 100$ and $h = 1$. Notice that the system remains stable for each  value of $p$ and for each $m  > 1$. Notice also that the predicted equilibria of the fluid model  agrees well with the predicted average steady state value of the agent based system, see Figure \ref{Fig: Equilibria}. }
\label{Fig: Comparison1}
\end{figure}

\begin{figure}
\includegraphics[width=\columnwidth]{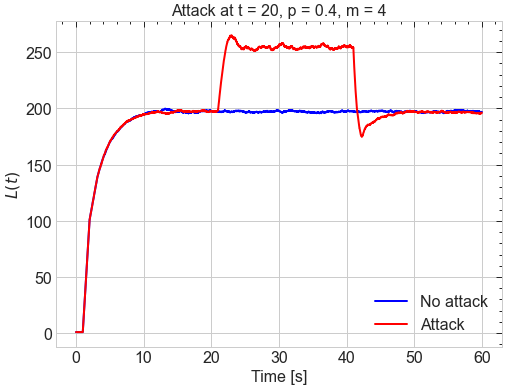}
\includegraphics[width=\columnwidth]{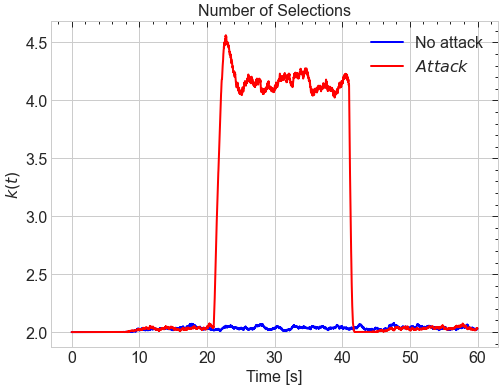}
\caption{Simulations of the Tangle with Algorithm \ref{alg: TMC} for $m =4$ and a Tips Inflation Attack with $p = 0.4$ starting at $t = 20$ and ending at $t = 40$. Each simulation shows {the average over} 100 realizations of the model with the time.}
\label{Fig: Comparison2}
\end{figure}

\begin{figure}
\includegraphics[width=\columnwidth]{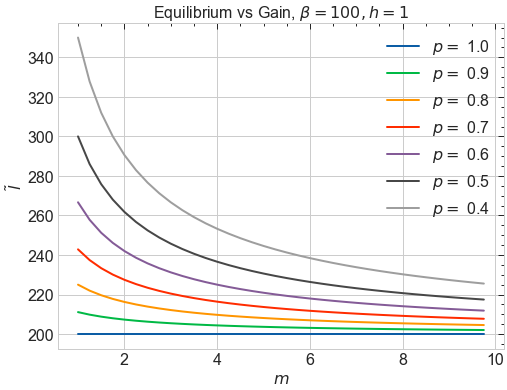}
\caption{Equilibria of the system with $\beta = 100$ and $h = 1$. The figure shows different values of the equilibrium $\tilde{l}$ for different values of $p$ and as the control parameter $m$ varies.}
\label{Fig: Equilibria}
\end{figure}

To showcase the performance, we consider the same agent based model of the Tangle, used in Section \ref{sec: validation}, with $\beta = 100$, $h = 1$ being subject to a Tips Inflation Attack  with $p = 0.4$ in two scenarios:
\begin{itemize}
    \item A continued attack from the beginning and for different values of $m$;
    \item A discontinuous attack, starting at $t = 20$ and ending at $t = 40$ with $m = 4$.
\end{itemize}
Figures \ref{Fig: Comparison1} and \ref{Fig: Comparison2} show the average  evolution of 100 different realizations of the Tangle for the two scenarios. Notice that, for all choices of $m$, the system remains stable and that the equilibria, predicted by \ref{eq: contrEquilibria} and shown in Figure \ref{Fig: Equilibria}, match the average steady state of the controlled agent based system. Furthermore, in the second scenario, the system reacts to the attack, rapidly adjusting the amount of tips selected by the honest nodes and manages to keep the system stable for the whole duration of the attack.\newline

One element that we believe it is worth stressing is that the controller does not reduce the effect of the attack to zero. While it might be possible to design a more complex controller (for instance including an integral action) to achieve this goal, this would require more computational power on the side of the honest nodes to compute the control action. While this is generally a trivial requirement under normal circumstances, in a DLT setting it is important to keep the amount of computations that honest nodes have to perform to as low as possible. Therefore, while the design of more complex controllers will be explored in a future work, we believe that the obtained ratio between performance and simplicity represents a good trade off, given the current setting.

\section{Conclusions and Future Lines of Research}
\label{sec: Conclusions}
In this paper we presented a fluid model of a class of DAG-based DLTs and designed a feedback control to reduce the effects of a Tips Inflation Attack. Our claims are backed by a thorough theoretical analysis, in the form of two stability theorems, and by extensive Monte Carlo simulations. The development of more complex control algorithms will be the focus of future research.

\section{Acknowledgements}

The authors wish to thank Daria Dziubałtowska for her help in the realization of this manuscript.

\bibliographystyle{IEEEtran}
\bibliography{references.bib}

\begin{thebibliography}{10}
\providecommand{\url}[1]{#1}
\csname url@samestyle\endcsname
\providecommand{\newblock}{\relax}
\providecommand{\bibinfo}[2]{#2}
\providecommand{\BIBentrySTDinterwordspacing}{\spaceskip=0pt\relax}
\providecommand{\BIBentryALTinterwordstretchfactor}{4}
\providecommand{\BIBentryALTinterwordspacing}{\spaceskip=\fontdimen2\font plus
\BIBentryALTinterwordstretchfactor\fontdimen3\font minus
  \fontdimen4\font\relax}
\providecommand{\BIBforeignlanguage}[2]{{%
\expandafter\ifx\csname l@#1\endcsname\relax
\typeout{** WARNING: IEEEtran.bst: No hyphenation pattern has been}%
\typeout{** loaded for the language `#1'. Using the pattern for}%
\typeout{** the default language instead.}%
\else
\language=\csname l@#1\endcsname
\fi
#2}}
\providecommand{\BIBdecl}{\relax}
\BIBdecl

\bibitem{kiayias2016blockchain}
A.~Kiayias, E.~Koutsoupias, M.~Kyropoulou, and Y.~Tselekounis, ``Blockchain
  mining games,'' in \emph{Proceedings of the 2016 ACM Conference on Economics
  and Computation}, 2016, pp. 365--382.

\bibitem{cullen2020resilience}
A.~Cullen, P.~Ferraro, C.~King, and R.~Shorten, ``On the resilience of
  dag-based distributed ledgers in iot applications,'' \emph{IEEE Internet of
  Things Journal}, vol.~7, no.~8, pp. 7112--7122, 2020.

\bibitem{Wang2020SoKDI}
Q.~Wang, J.~Yu, S.~Chen, and Y.~Xiang, ``Sok: Diving into dag-based blockchain
  systems,'' \emph{ArXiv}, vol. abs/2012.06128, 2020.

\bibitem{overko2020spatial}
R.~Overko, R.~Ord{\'o}nez-Hurtado, S.~Zhuk, P.~Ferraro, A.~Cullen, and
  R.~Shorten, ``Spatial positioning token (sptoken) for smart mobility,''
  \emph{IEEE Transactions on Intelligent Transportation Systems}, 2020.

\bibitem{katsikouli2020distributed}
P.~Katsikouli, P.~Ferraro, H.~Richardson, H.~Cheng, S.~Anderson, D.~Mallya,
  D.~Timoney, M.~Masen, and R.~Shorten, ``Distributed ledger enabled control of
  tyre induced particulate matter in smart cities,'' \emph{Frontiers in
  Sustainable Cities}, p.~48, 2020.

\bibitem{ferraro2021personalised}
P.~Ferraro, L.~Zhao, C.~King, and R.~Shorten, ``Personalised feedback control,
  social contracts, and compliance strategies for ensembles,'' \emph{arXiv
  preprint arXiv:2103.07261}, 2021.

\bibitem{popov2015}
S.~{Popov}, ``{The Tangle},'' 2015.

\bibitem{ferraro2018distributed}
P.~Ferraro, C.~King, and R.~Shorten, ``Distributed ledger technology for smart
  cities, the sharing economy, and social compliance,'' \emph{Ieee Access},
  vol.~6, pp. 62\,728--62\,746, 2018.

\bibitem{ferraro2019stability}
------, ``On the stability of unverified transactions in a dag-based
  distributed ledger,'' \emph{IEEE Transactions on Automatic Control}, vol.~65,
  no.~9, pp. 3772--3783, 2019.

\bibitem{karame2016bitcoin}
G.~O. Karame and E.~Androulaki, \emph{Bitcoin and blockchain security}.\hskip
  1em plus 0.5em minus 0.4em\relax Artech House, 2016.

\bibitem{gervais2016security}
A.~Gervais, G.~O. Karame, K.~W{\"u}st, V.~Glykantzis, H.~Ritzdorf, and
  S.~Capkun, ``On the security and performance of proof of work blockchains,''
  in \emph{Proceedings of the 2016 ACM SIGSAC conference on computer and
  communications security}, 2016, pp. 3--16.

\bibitem{Kusmierz2019}
B.~Kusmierz, W.~Sanders, A.~Penzkofer, A.~Capossele, and A.~Gal, ``{Properties
  of the Tangle for Uniform Random and Random Walk Tip Selection},'' in
  \emph{2019 IEEE International Conference on Blockchain (Blockchain)}, 2019,
  pp. 228--236.

\bibitem{popov2019equilibria}
S.~Popov, O.~Saa, and P.~Finardi, ``Equilibria in the tangle,'' \emph{Computers
  \& Industrial Engineering}, vol. 136, pp. 160--172, 2019.

\bibitem{2020Muller}
S.~M{\"u}ller, A.~Penzkofer, B.~Ku{\'{s}}mierz, D.~Camargo, and W.~J. Buchanan,
  ``Fast probabilistic consensus with weighted votes,'' in \emph{Proceedings of
  the Future Technologies Conference (FTC) 2020, Volume 2}, K.~Arai, S.~Kapoor,
  and R.~Bhatia, Eds.\hskip 1em plus 0.5em minus 0.4em\relax Cham: Springer
  International Publishing, 2021, pp. 360--378.

\bibitem{cullen2021}
A.~Cullen, P.~Ferraro, W.~Sanders, L.~Vigneri, and R.~Shorten, ``{Access
  Control for Distributed Ledgers in the Internet of Things: A Networking
  Approach},'' \emph{IEEE Internet of Things Journal}, pp. 1--1, 2021.

\end{thebibliography}

\end{document}